\documentstyle[aps,epsf]{revtex}
\begin{document}
\draft
\title{Long-time behavior of MHD shell models}
\author{P. Frick$^{1}$, G. Boffetta$^{2}$, P. Giuliani$^{3}$,
S. Lozhkin$^{1}$ and D. Sokoloff$^{1}$}
\address{
        $^{1}$ Institute of Continuous Media Mechanics,
         Korolyov 1, 614061 Perm, Russia\\
        $^{2}$ Dipartimento di Fisica Generale and INFM,\\
         Universit\`a di Torino, Via P. Giuria 1, 10125 Torino, Italy\\
        $^{3}$ Dipartimento di Fisica, Universit\`a della Calabria,
         87036 Rende (Cs), Italy}
\maketitle
\begin{abstract}
The long time behavior of velocity-magnetic field alignment
is numerically investigated in the framework of MHD shell model.
In the stationary forced case, the correlation parameter $C$
displays a nontrivial behavior with long periods of high
variability which alternates with  periods of almost constant $C$.
The temporal statistics of correlation is shown
to be non Poissonian, and the pdf of constant sign periods displays
clear power law tails. 
The possible relevance of the model for geomagnetic dynamo problem
is discussed.
\end{abstract}

\vspace{0.5cm}
It is  generally believed  that the turbulent motions of an electrically
conducting fluid or plasma can excite turbulent, fluctuating magnetic fields,
the so-called turbulent dynamo.
Traditionally the analysis of
turbulent dynamo is started with the kinematic stage where the magnetic field is
supposed to be weak enough to neglect its influence on the fluid velocity
field, which is considered as a prescribed random field with
time-independent statistical properties.
The response of the magnetic field to the fluid motion results in a
saturation of the magnetic field growth. In the following one can expect 
the establishment of a kind of  statistical equilibrium.
The aim of this paper is to investigate numerically an extremely long time
behavior of the nonlinear dynamics.

The possibility to follow the long-term evolution of the turbulent 
magnetic field by direct numerical simulations is very limited, because
of the required large magnetic and kinematic Reynolds numbers 
(see e.g. \cite{DNS}).
That is why our discussion is based on the shell model of small scale
MHD turbulence introduced in \cite{FS},
which reaches a statistically stationary state with a
spectral index depending on the level of cross-helicity and magnetic
helicity.

The basic idea of any shell model of fully developed turbulence
is to retain either one real or complex mode
(in our case this is  $U_n$ for velocity and $B_n$ for magnetic field)
as a  representative of all
corresponding modes in the shell with wave number $k_n<|\vec{k}|<k_{n+1}$,
$k_n=2^n$, and to introduce a set of ODE, which mimics the
original nonlinear PDE (i.e. Navier-Stokes equations in the pure
hydrodynamical case). For an introduction to shell models the readers 
are referred to \cite{BJPV98}.

 Here we will use the MHD-shell model  introduced in \cite{FS} and rewritten
 for  3D turbulence as
\begin{eqnarray}
(d_t + Re^{-1} k_n^2)U_n
&=& i k_n \Bigl \lbrace (U_{n+1}^*U_{n+2}^*
- B_{n+1}^*B_{n+2}^* )
- \frac{1}{4} (U_{n-1}^*U_{n+1}^* - B_{n-1}^*B_{n+1}^* ) \nonumber \\
&-& \frac{1}{8} (U_{n-2}^*U_{n-1}^*
- B_{n-2}^*B_{n-1}^* )\Bigl \rbrace +  f_n. 
\label{goy_u}
\end{eqnarray}
\begin{eqnarray}
(d_t + Rm^{-1} k_n^2)B_n
&=& \frac{i k_n}{6} \Bigl \lbrace
(U_{n+1}^*B_{n+2}^* - B_{n+1}^*U_{n+2}^* )
+ (U_{n-1}^*B_{n+1}^* - B_{n-1}^*U_{n+1}^* ) \nonumber \\
&+&  (U_{n-2}^*B_{n-1}^* - B_{n-2}^*U_{n-1}^* )\Bigl \rbrace.
\label{goy_b}
\end{eqnarray}
Here $Re$ is the Reynolds number, $Rm = Re\cdot Pr_m$ is the magnetic
Reynolds number, and $Pr_m$ is the magnetic Prandtl number.
$f_n$ is the external forcing which acts on the velocity field only.
If $B_n = 0$,  one obtains the so-called GOY shell model \cite{BJPV98,OY89,FDB},
widely used for the Navier-Stokes equations

 In the limit $Re,Rm\to \infty$, equations (\ref{goy_u},\ref{goy_b})
conserve three quadratic quantities
corresponding to the three quadratic invariants of  inviscid
MHD flows: total energy 
$E=E_V+E_B = \sum_{n} |U_n|^2 + |B_n|^2$, 
cross helicity 
$H_C = \sum_{n} Re(U_n B_n^{*})$ and
magnetic helicity 
$H_B = \sum_{n} (-1)^{n} k_{n}^{-1} |B_n|^2$ 
\cite{Biskamp93}.
To proceed further  let us note that the nonlinear terms
of (\ref{goy_u},\ref{goy_b}) identically vanish for $U_n=\pm B_n$
(Alfv\'enic fixed points).

By considering an initial distribution of kinetic energy
with a weak magnetic energy ($E_B \ll E_V$), and setting $f_n=0$, 
one arrives at the dynamo problem in free-decaying turbulence.
This approach is a reasonable approximation, because
the time of energy decay is much longer than the characteristic time
of the magnetic field growth.

The dynamics of the decaying case has been investigated in detail
in \cite{FS}. Starting from a very low level, 
 the magnetic energy reaches
about $1/10$ of kinetic energy in about one turnover time.
Then appears a relatively long period of nonlinear evolution
(order of $20-40$ turnover times) after which the ratio $E_B/E_V$
remains constant (and of order $1$).
The spectral index of magnetic and velocity fields is usually close to
the classical value $-5/3$ with a constant energy flux over a wide spectral
range. It should be noted that the cross-helicity
$H_C$ remains close to zero during the whole decay.

An essentially different picture is observed in the
forced turbulence sustained by a constant external force
$f_0$ acting on one shell only ($n=0$).
Starting again with a  weak magnetic field, the system at the initial evolution
stage displays the same behavior as in the free decaying case, and
a statistically stable state (with Kolmogorov scaling)
seems to be established at time $t\sim 10$ \cite{FS}.
However, longer simulations under similar forcing \cite{GC}
revealed that after relatively long evolution  this state is
replaced by another one. In contrast to the initial stage of evolution,
the magnetic field is strongly correlated (or anticorrelated) with
the velocity field.
The value of cross-helicity $H_C$ is close to its maximal value
(i.e. the correlation parameter, $C=2H_C/E$, is close to $1$ or to $-1$).
Let us note that the states with high correlation between the magnetic and
velocity fields are well known in MHD (e.g. Alfv\'en waves).
From dynamical viewpoint, high correlations imply strong
depletion of the nonlinear terms in (\ref{goy_u},\ref{goy_b}) and thus a
weak energy flux. As a consequence, the slope of  the spectral index is expected
to be very steep, as observed  in  numerical simulations \cite{Cracow}.

\begin{figure}[htb]
\centerline{
\epsfxsize=8truecm
\epsfbox{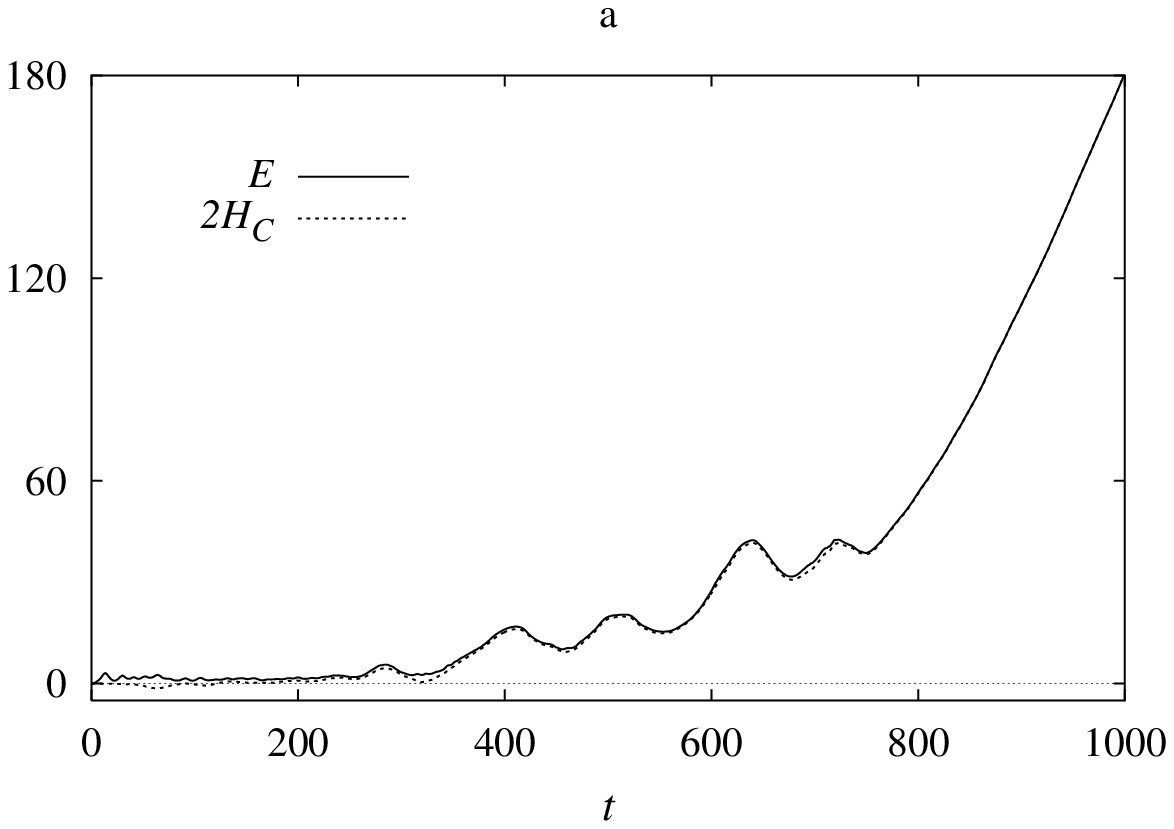},
\epsfxsize=8truecm
\epsfbox{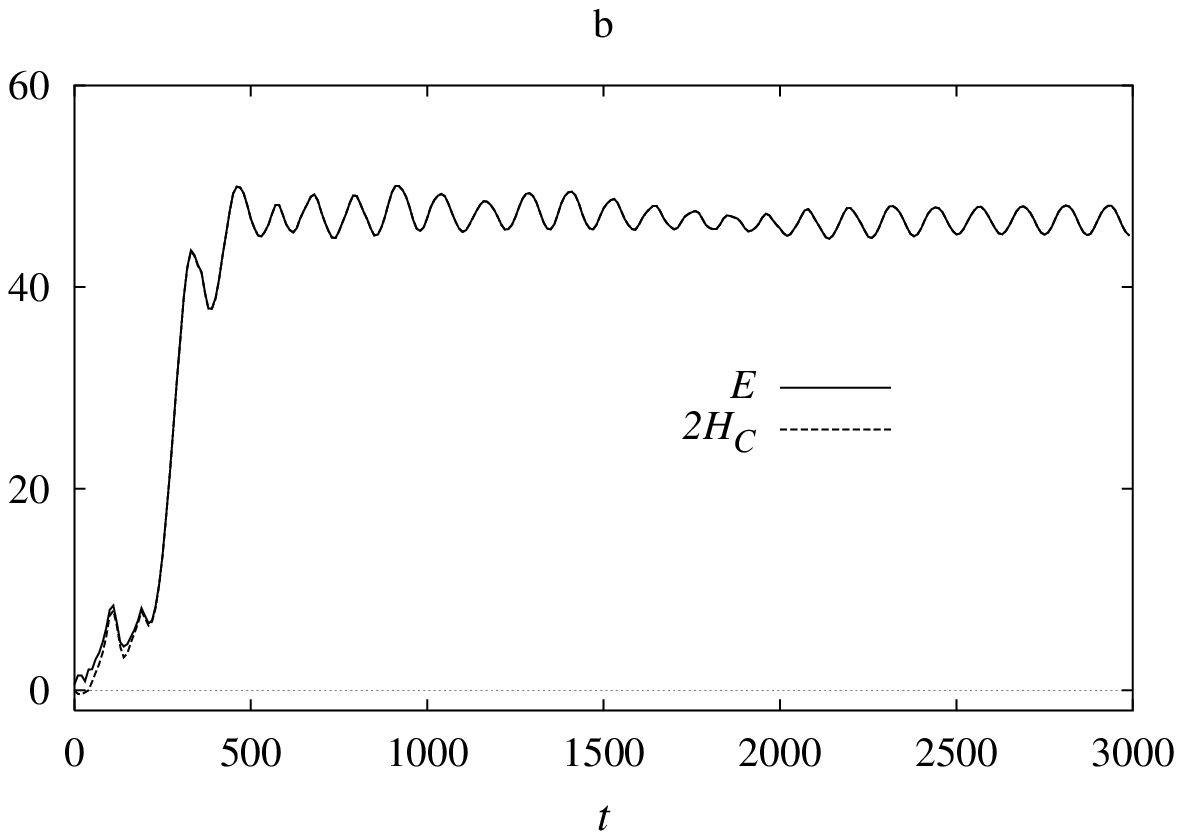}
}
\caption{Time evolution of total energy $E$
and cross helicity $H_{C}$ for the MHD shell model
with constant forcing and slightly different initial conditions. 
In both cases we observe strong correlation between kinetic and
magnetic variables which reduces the energy flux.}
\label{fig1}
\end{figure}

The specific behavior of a given solution of the shell model
depends on the choice of initial conditions.
The difference between two evolution
tracks under slightly varied initial conditions can be surprisingly great.
To illustrate this point,  Figure~\ref{fig1} gives the evolution
of two simulations  under slightly  different initial conditions.
At long times we observed either an unlimited growth of magnetic and
kinetic energy (as in Figure~\ref{fig1}a), or a very long oscillatory
behavior (as shown  in  Figure~\ref{fig1}b) \cite{Cracow}.
The dependence of the observed correlated state on
the forcing  has been discussed by Giuliani \cite{Giul} who replaced the
constant force by the Gaussian random forcing exponentially correlated in
time.
The system then  moves randomly  between  the correlated and
anticorrelated states. This behavior assures stationarity
with a well defined mean energy flux to  small scales.

The long-term evolution of MHD shell model presented in Figure~\ref{fig1}
demonstrates a drastic variation in the total energy of the system,
suggesting a strong inflow or outflow of the energy.
Since our main focus is a basically isolated system, we
provide conservation of the kinetic energy in the largest scale ($n=0$)
by applying a different kind of force.
At each time step, we force the value of $|U_0|$ to a given
constant, while the phase is left free to follow the dynamics.

\begin{figure}[htb]
\centerline{
\epsfxsize=8truecm
\epsfbox{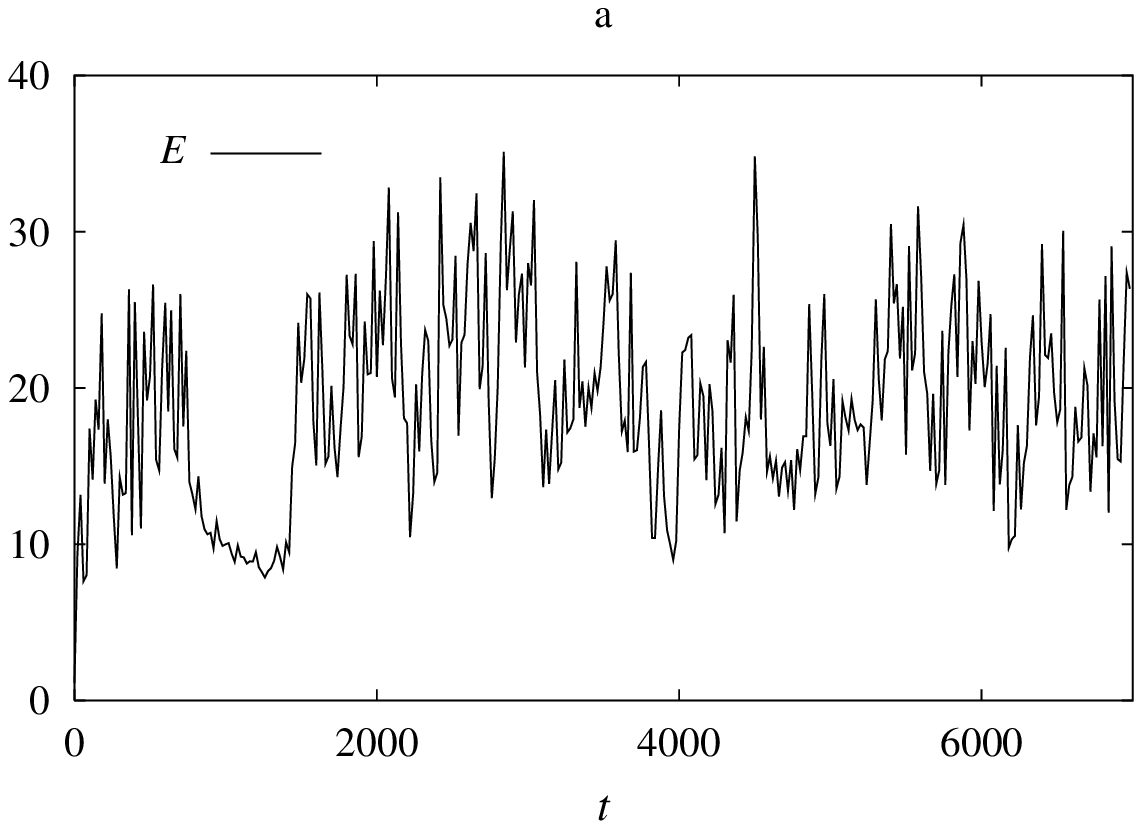}
\epsfxsize=8truecm
\epsfbox{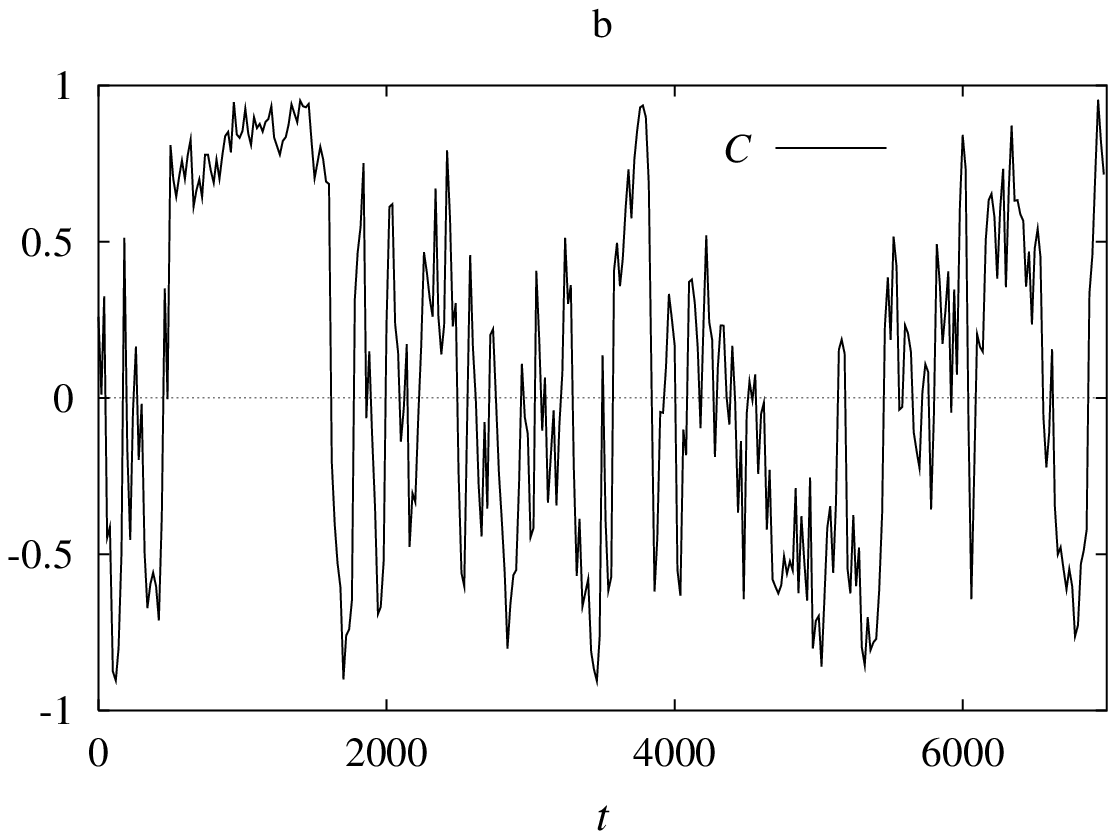}}
\caption{Time evolution of (a) total energy $E$ and
cross helicity $H_{C}$ and (b) the correlation parameter $C$
in the ``isolated'' MHD shell model. 
The kinetic energy of the first shell $|U_0|^2$ is kept constant by 
rescaling the amplitude of $U_0$ at every time step.}
\label{fig2}
\end{figure}

With this kind of forcing  most of the time
the kinetic and magnetic energy oscillate around the mean value
with the small correlation parameter $C$ (Figure~\ref{fig2}).
However, one can observe a  relatively long stage of high correlation
and small energy oscillation  (i.e. for $500<t<1500$
in Figure~\ref{fig2}).

\begin{figure}[htb]
\centerline{
\epsfxsize=12truecm
\epsfbox{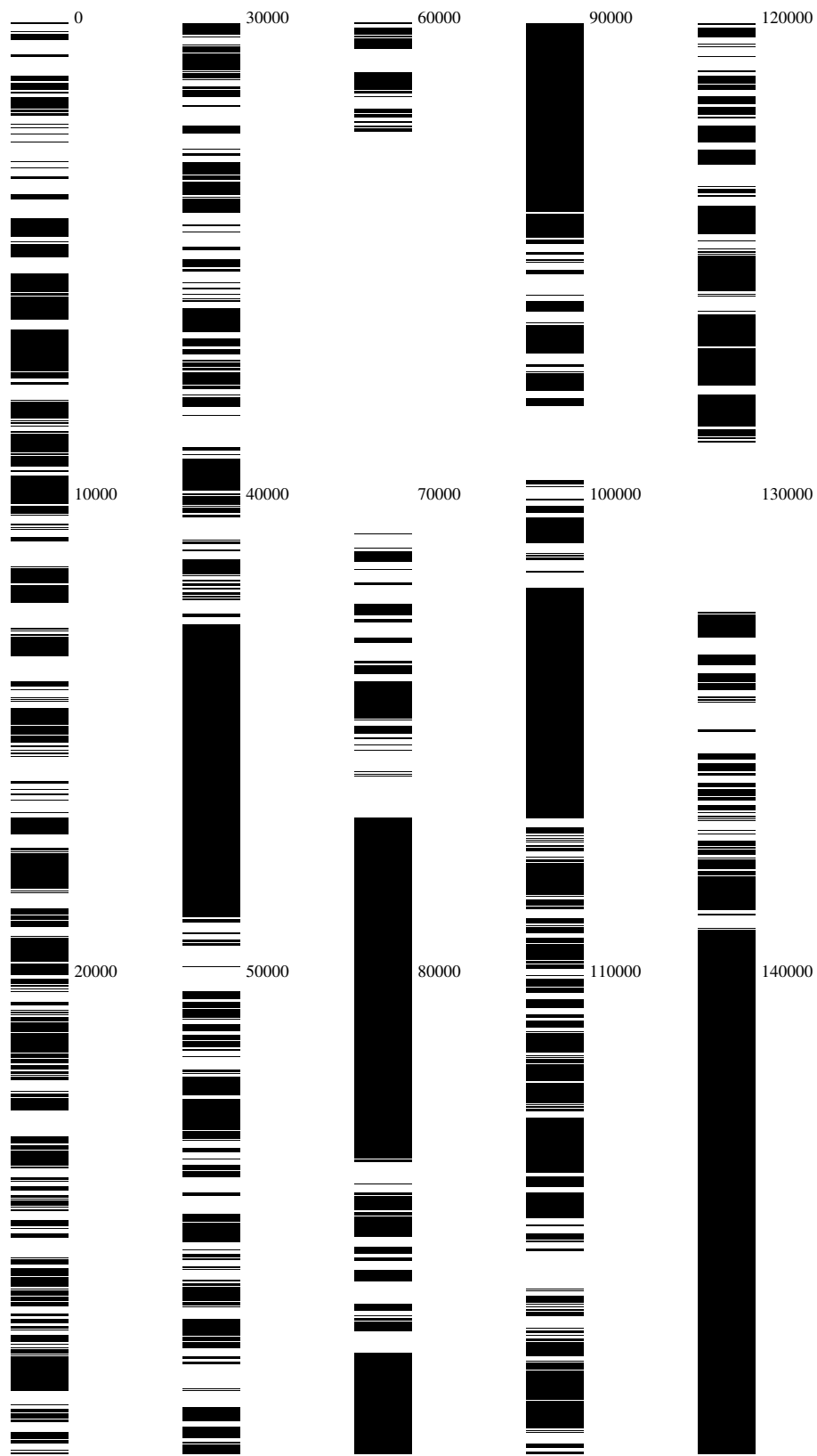}}
\caption{Sequence of time reversal for the correlation parameter
$C$ in the long shell model simulation.}
\label{fig3}
\end{figure}

We performed simulations of the isolated model for a very long period 
of time up to $t = 150000$.
During the evolution, the alignment $C$ changed the sign $2832$ times.
The sequence of the sign  reversals is shown in Figure~\ref{fig3}
as a sequence of black ($C>0$) and white ($C<0$) strips.
This type of visualization is chosen by analogy with
the geomagnetic studies where different polarities of the geomagnetic field
is usually represented by white and black strips  (see e.g. \cite{AS}).

\begin{figure}[htb]
\centerline{
\epsfxsize=12truecm
\epsfbox{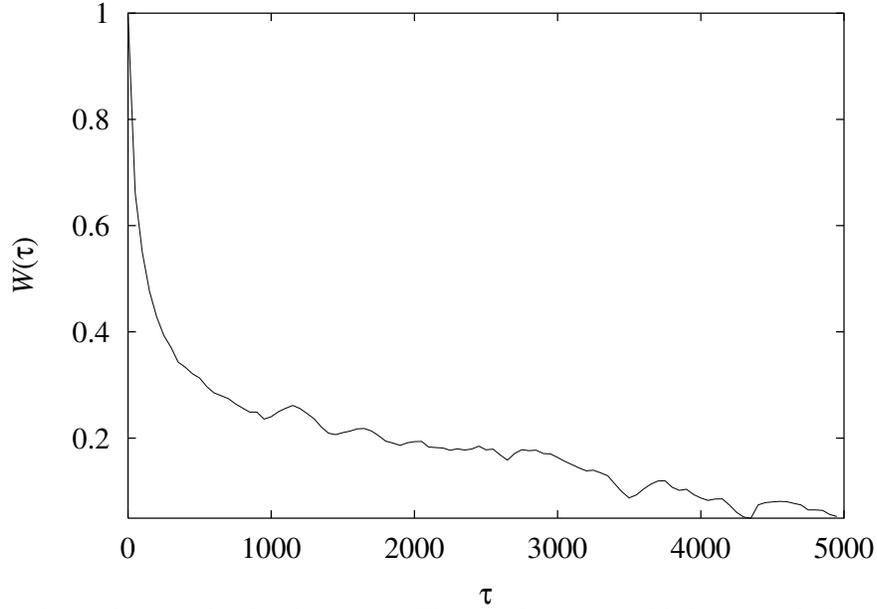}}
\caption{Temporal correlation function for the alignments. The initial
exponential behavior with characteristic time $\tau_0\simeq 200$ is
clearly observable.}
\label{fig4}
\end{figure}

Alternation  of black and white strips seems to be  a random process
which can be investigated by statistical tools.
We computed the correlation function of alignments by introducing
a stepwise function $f(t) = 1$ if $C(t) > 0$ and $f(t) = -1$
if $C(t) \leq 0$.
Figure~\ref{fig4} shows the time correlation defined by
\begin{equation}
W (\tau) = {{\int_0^{T-\tau} (f(t) f(t+\tau)-\langle f \rangle ^2)d t} \over
{(T-\tau)
\langle (f(t)-\langle f \rangle )^2 \rangle}},
\end{equation}
where $\langle f \rangle$ represents the time average and $T=150000$ is the total length
of the simulation.
We can expect {\it a priori} that the estimate of $W$ is possible up to values of $\tau$
not larger than $\tau \approx 5000$
(it corresponds to $T/\tau \approx 30$). In practice, $W$ becomes noisy
even earlier, at $\tau = 1000$.
A quasi-exponential decay of correlation,
$W(\tau) \sim \exp(-\tau/\tau_0)$ with characteristic
time scale $\tau_0=200$ is observable for $\tau < 500$.
This time scale is in crude agreement with mean value of the epoch of a given
sign of alignment. In the other words, the memory of a given epoch is
preserved in a few following epoch at most.

\begin{figure}[htb]
\centerline{
\epsfxsize=12truecm
\epsfbox{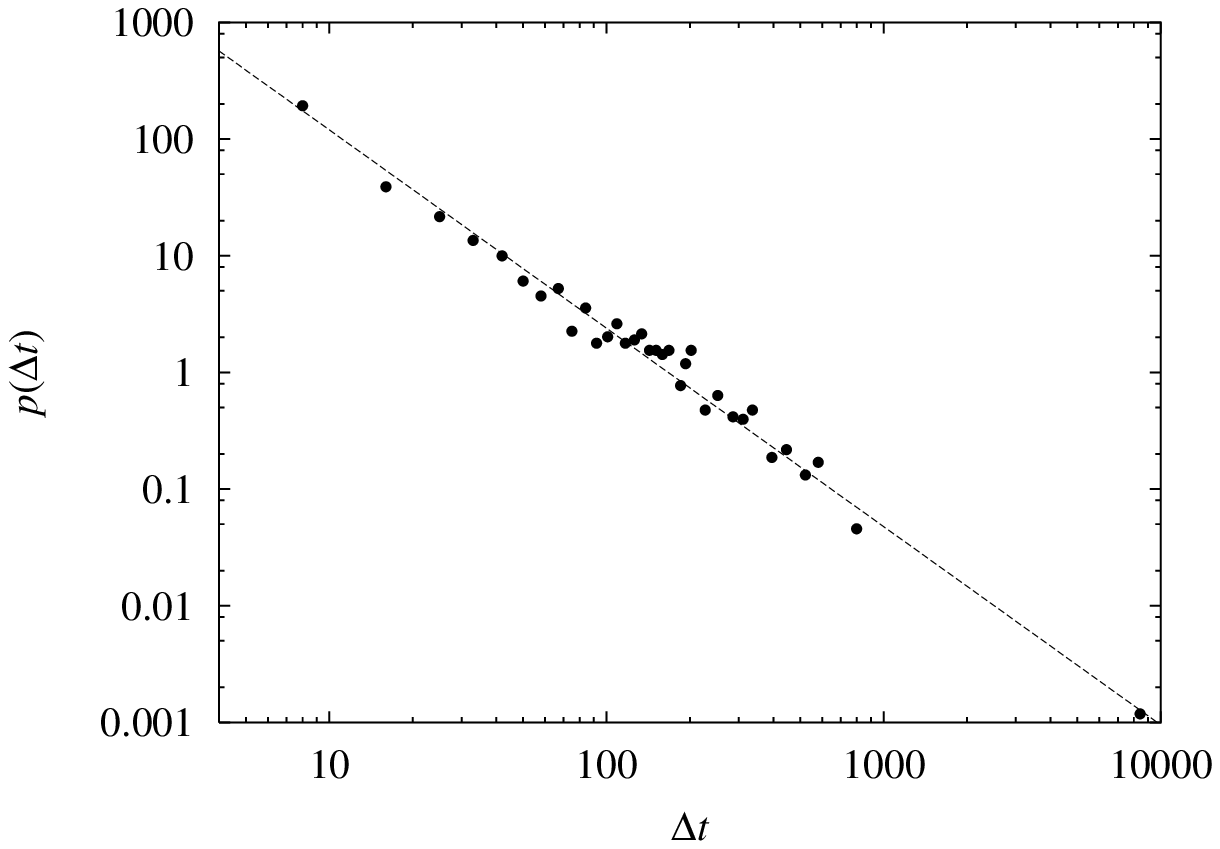}}
\caption{Probability density distribution of the delays $\Delta t$
between successive alignment inversions in Figure~\ref{fig3}. The
clear power law scaling reveals the non--Poissonian nature of 
the process. The dashed line is the fit $p(\Delta t)\simeq \Delta t^{-1.7}$}
\label{fig5}
\end{figure}

The intervals of steady sign of alignment vary from $\Delta t = 1$ (determined
by the numerical resolution) up to $\Delta t = 8 421$. This high variability of
$\Delta t$ is hardly compatible with Poissonian nature of epoch alternation. 
Figure~\ref{fig5}, presenting the plot of the probability density
$p(\Delta t)$, supports this statement. We observe a clear power 
law behavior
$p(\Delta t) \simeq \Delta t^{-1.7}$ for a wide range of $\Delta t$.
Let us note that non Poissonian nature of temporal statistics
in MHD shell model has been already observed in a different
context \cite{BCGVV99}.


In this letter we have investigated numerically
the statistics of a very long time evolution of
MHD shell model. In the stationary forced case
the kinetic and magnetic energy exhibit a strong
chaoticity with intermediate periods of small
variability. 
The correlation $C$ between the velocity and
magnetic field reflects this intermittent behavior. In the period of 
strong variability
$C$ fluctuates very rapidly, while it sticks
close to the maximum value $C=\pm 1$ in the
low variability periods.

It can be interesting to compare our findings on the long-term
behavior of the MHD shell model with low-order models of nonlinear
dynamo.
For example, the well known Rikitake model \cite{R58} indicates chaotic 
reversals of the magnetic field, but the
absolute values of the magnetic field between reversals are close.
More complex models of geodynamo provide the chaotic sequence
of reversals with the chaotic evolution of magnetic energy
between reversals and the fractal distribution of reversal instants
(\cite{AS} and references therein).

\vspace{0.5cm}
This work was partly supported by RFBR under grants 99-01-00362a, 
00-05--64062a and by CNR Special Project 
(``Fully developed turbulence in plasmas'').


\newpage

\begin{thebibliography}{99}

\bibitem{DNS}
M.J. Korpi, A. Brandenburg, A. Shukurov, I. Tuominen and \AA. Nordlund,
Astrophys. J. Lett {\bf 514} (1999) L99.

\bibitem{FS}
P. Frick and D. Sokoloff,
Phys. Rev. E {\bf 57} (1998) 4195.

\bibitem{BJPV98}
T. Bohr, M. Jensen, G. Paladin and A. Vulpiani,
{\it Dynamical Systems Approach to Turbulence}
(Cambridge University Press, Cambridge)
(1998).

\bibitem{OY89}
K. Ohkitani and M. Yamada,
Prog. Theor. Phys.  {\bf 81} (1989) 329.

\bibitem{FDB}
P. Frick, B. Dubrulle and A. Babiano,
Phys. Rev. E {\bf 51} (1995) 5582.

\bibitem{Biskamp93}
D. Biskamp,
{\it Nonlinear Magnetohydrodynamics}
Cambridge University Press, Cambridge
(1993).

\bibitem{GC}
P. Giuliani and V. Carbone,
Europhys. Lett.  {\bf 43} (1998) 527.

\bibitem{Cracow}
P. Frick, S. Lozhkin and D. Sokoloff,
in {\it Plasma Turbulence and Energetic Particles in Astrophysics}
edited by M. Ostrowski and R. Schlikeiser
(Krak\'ow) (1999) 190.

\bibitem{Giul}
P. Giuliani,
in {\it Non-linear MHD waves and turbulence}
edited by T. Passot and P.L. Sulem
(Springer Verlag) (1999) 331.

\bibitem{AS}
A. Anufriev and D. Sokoloff,
Geophys. Astrophys. Fluid Dynamics {\bf 74} (1994) 207.

\bibitem{BCGVV99}
G. Boffetta, V. Carbone, P. Giuliani, P. Veltri and A. Vulpiani,
Phys. Rev. Lett.  {\bf 83} (1999) 4662.

\bibitem{R58}
T. Rikitake,
Proc. Camb. Phil. Soc.  {\bf 54} (1958) 89.

\end{thebibliography}
\end{document}